\documentclass[%
 aip,
 amsmath,amssymb,
 reprint,%
]{revtex4-1}

\usepackage{graphicx}
\usepackage{dcolumn}
\usepackage{bm}

\usepackage[utf8]{inputenc}
\usepackage[T1]{fontenc}
\usepackage{mathptmx}
\usepackage{comment}
\usepackage{makecell}
\usepackage{textcomp}
\usepackage{hyperref}
\usepackage[]{lineno}

\modulolinenumbers[5]

\begin{document}

\preprint{AIP/123-QED}

\title{Feasibility study of a compact Neutron Resonance Transmission Analysis instrument}


\author{Ezra M. Engel}
\affiliation{Department of Nuclear Science and Engineering, Massachusetts Institute of Technology, Cambridge, MA 02139}
\author{Ethan A. Klein}
\affiliation{Department of Nuclear Science and Engineering, Massachusetts Institute of Technology, Cambridge, MA 02139}
\author{Areg Danagoulian}
\altaffiliation[Corresponding author: ]{aregjan@mit.edu}
\affiliation{Department of Nuclear Science and Engineering, Massachusetts Institute of Technology, Cambridge, MA 02139}

\date{\today}

\begin{abstract}
Neutron Resonance Transmission Analysis (NRTA) uses resonant absorption of neutrons to infer the absolute isotopic composition of a target object, enabling applications in a broad range of fields such as archaeology, enrichment analysis of nuclear fuel, and arms control treaty verification.  In the past, NRTA involved large user facilities and complex detector systems. However, recent advances in the intensity of compact neutron sources have made compact neutron imaging designs increasingly feasible. This work describes the Monte Carlo (MC) based design of a compact epithermal NRTA radiographic instrument which uses a moderated, compact deuterium-tritium (DT) neutron source and an epithermal neutron detector. Such an instrument would have a wide range of applications, and would be especially impactful for such scenarios as nuclear inspection and arms control verification exercises, where system complexity and mobility may be of critical importance. The MC simulations presented in this work demonstrate accurate time-of-flight (TOF) reconstructions for transmitted neutron energies, capable of differentiating isotopic compositions of nuclear material with high levels of accuracy. A new generation of miniaturized and increasingly more intense neutron sources will allow this technique to achieve measurements with greater precision and speed, with significant impact on a variety of engineering and societal problems.
\end{abstract}

\maketitle


\section{Introduction} \label{S:1}

Many high- and mid-Z nuclei exhibit neutron-induced resonances in the epithermal range (from 1 eV to 100 eV).  Neutron Resonance Transmission Analysis (NRTA) uses the isotopic uniqueness of these resonances in a transmission measurement mode to determine isotopic and elemental compositions of target materials.  Depending on the isotope, most of these interactions are the result of either $(n,\gamma)$, $(n,n')$, or $(n,$fission$)$ processes.  The transmitted spectrum exhibits attenuation dips at energies that uniquely map to the resonance energies of the nuclei in the target material.  
In these measurements, a pulsed neutron beam (e.g. via spallation by a proton beam) is used in combination with a detector sensitive to epithermal neutrons, such as a lithium doped scintillator or a boron doped microchannel plate (MCP)~\cite{ref:tremsin}.  The time between the neutron generation pulse and the time at which the neutron is detected provides the TOF of individual neutrons, enabling a reconstruction of their energy and thus achieving spectroscopic radiography.  In combination with known cross section data, the transmission spectrum can be used to precisely and simultaneously determine the identity and the areal density of the multiple isotopes that make up the target object.  A pixelated detector can further provide two-dimensional coordinate information~\cite{ref:losko,ref:tremsin,vogel2017neutron}, which can then be used to determine the two-dimensional areal density of the isotopes, thus producing isotopic imaging of the target.

Other neutron transmission techniques exist as well.  In particular, work by~\citeauthor{ref:blackburn2007fast}\cite{ref:blackburn2007fast} and ~\citeauthor{perticone2019fast}~\cite{perticone2019fast} has shown that $\sim$MeV fast neutrons can be used to map the composition of a number of low-Z elements, such as carbon, oxygen, and nitrogen.   Most mid-Z to high-Z elements however do not exhibit experimentally distinguishable resonances at $\sim$MeV energies, however.  Other radiographic methods are available to determine the hydrogenous content of an object~\cite{rahon2016spectroscopic,rahon2020hydrogenous,sowerby}.  Additionally, other applications of epithermal neutrons exist.  In particular, Boron Neutron Capture Therapy uses intense fluxes of epithermal neutrons in oncological applications, as described in detail by \citeauthor{doi:harling}\cite{doi:harling} and \citeauthor{forton2009overview}\cite{forton2009overview}.

In recent years NRTA has been used in many areas of nuclear engineering~\cite{ref:chichester2012assessing,ref:chichester2012JNMM,bourke2016non} and archaeology~\cite{schillebeeckx2015neutron,andreani2009novel}. 
Epithermal NRTA has also been proposed as a platform for warhead verification in arms control treaties~\cite{ref:hecla2018nuclear,ref:engel2019}. Thus the isotopic specificity of NRTA gives it a very broad range of applicability. For historical and practical reasons, NRTA experiments have primarily taken place at facilities which were designed and built for nuclear data measurements.  Most of these facilities possess kilometer-long accelerators, significantly limiting the application of an otherwise powerful technique. The utility of NRTA can thus be significantly extended by the design of compact, possibly mobile, and relatively low cost platforms. Pulsed deuterium-tritium (DT) neutron sources, augmented by carefully optimized moderation, shielding, and collimation, can produce a compact and precise beam, suitable for mobile applications.  When combined with a shielded neutron detector, such a beam can constitute a novel instrument for isotope-specific radiography of objects composed of mid- and high-Z elements. In this Monte Carlo-based study we focus on the optimization of compact DT-based platforms, showing the feasibility and practicality of such platforms when it comes to materials analysis.  This technique can in the future benefit from ongoing research in high intensity, fast-pulsed DT and deuterium-deuterium (DD) sources~\cite{podpaly2018environment,ref:starfire}.  In this work, we perform Monte Carlo based feasibility and optimization studies of such an instrument, presenting an optimal configuration for high-Z isotopic imaging and demonstrating the feasibility of such a platform for mobile nuclear security applications. While recent work was performed on the development of compact thermal and epithermal sources~\cite{hasemi2017optimization, paradela2016neutron}, no design studies have been performed on the development of a compact and complete NRTA system.

\section{Monte Carlo Simulations} \label{S:2}
Monte Carlo simulations were performed to model two effects: the impact of moderation process on the accuracy of the TOF energy reconstruction; the spectral sensitivity to changes in isotopic composition of the target.  The purpose of the simulations was to independently model shielding configuration and moderator thickness and to optimize these criteria.  Optimization of the moderator improved the neutron epithermalization efficiency and optimization of the shielding geometry improved the TOF energy reconstruction and reduced the background.

\subsection{Simulations}
\begin{figure}[h]
\centering\includegraphics[width=0.8\linewidth]{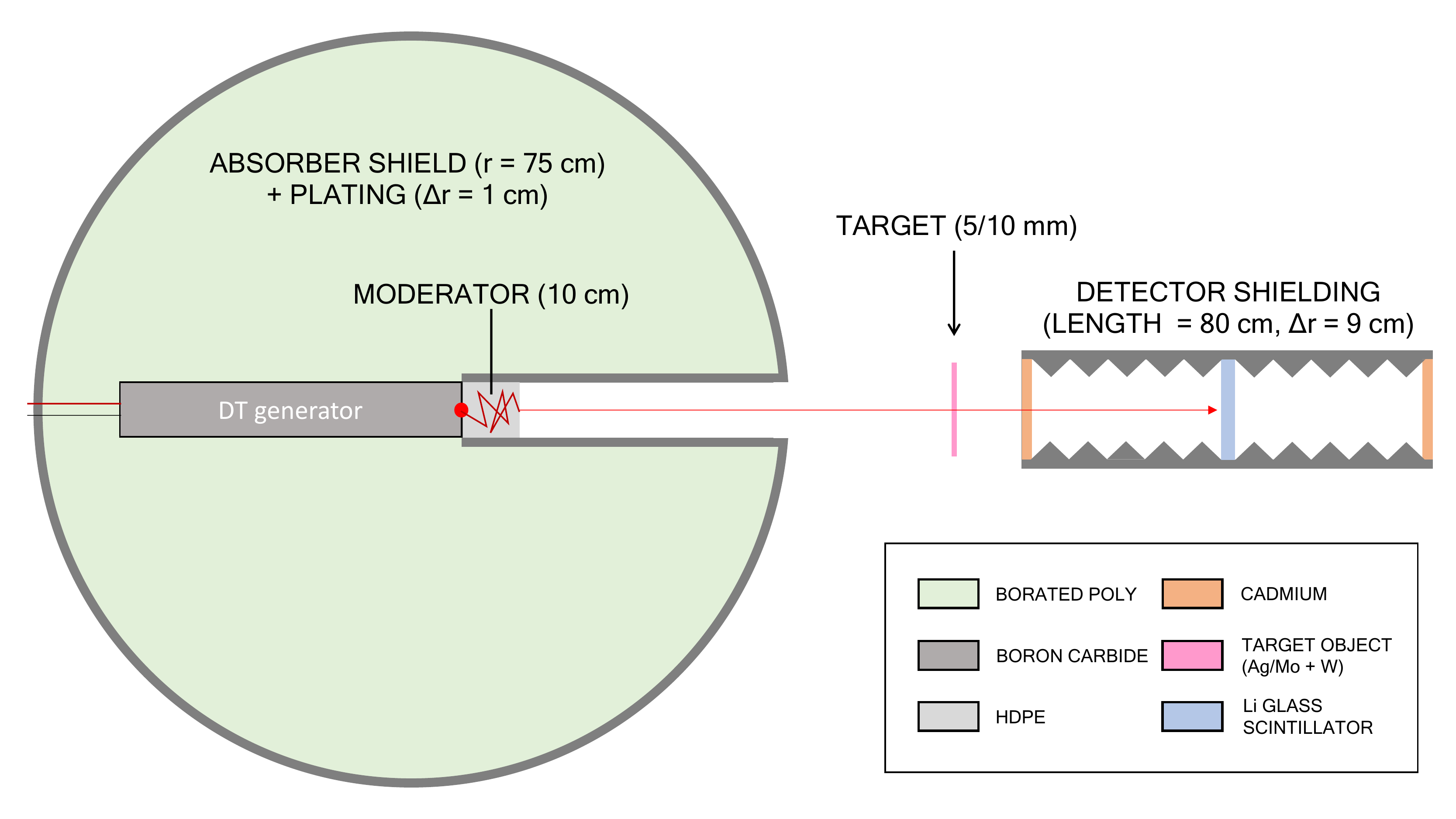}
\centering\includegraphics[angle=270,width=0.8\linewidth]{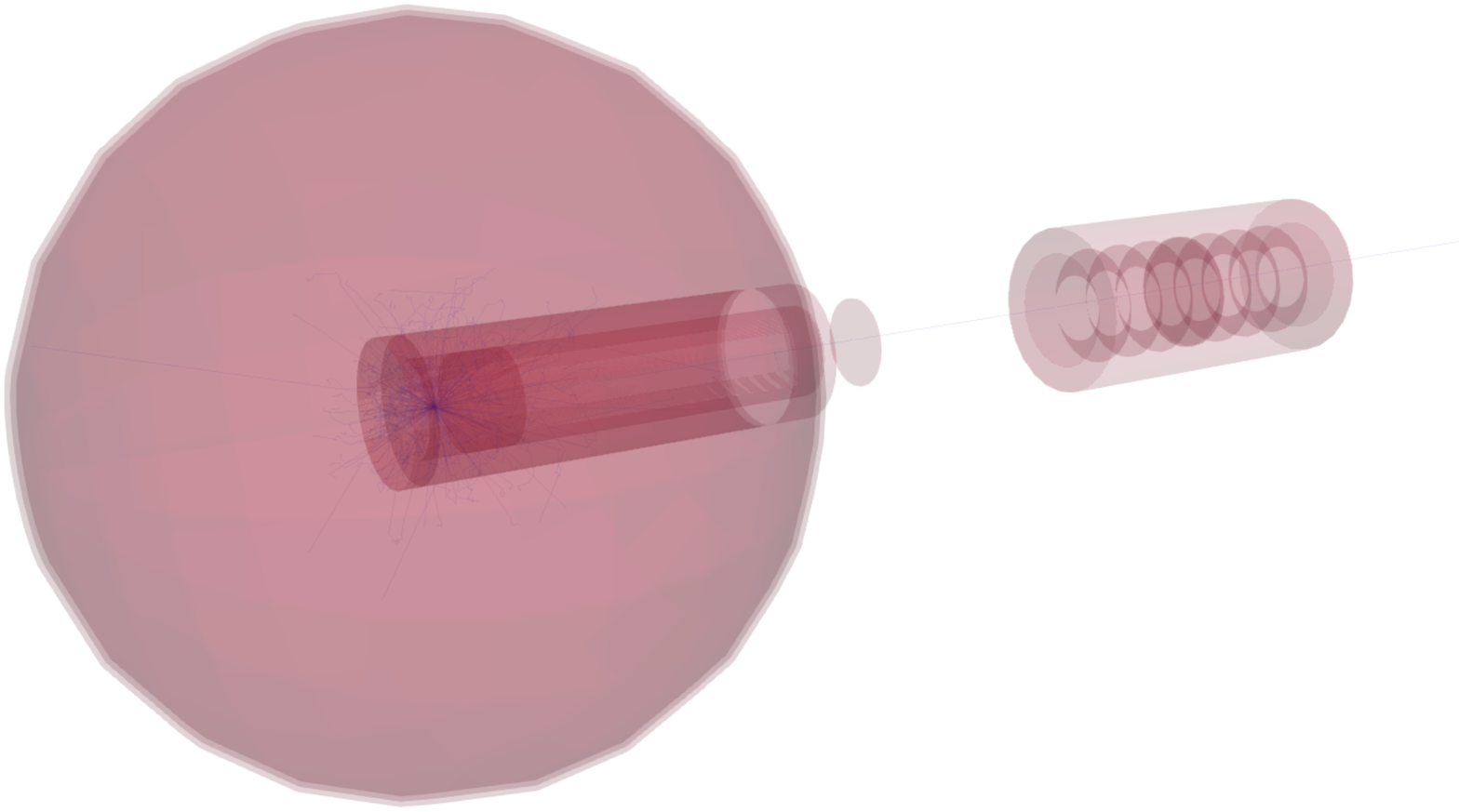}
\caption{Top: schematic of a compact NRTA platform using a DT source (not to scale).  Bottom: a Geant4 rendering of the simulation.  Due to its insignificant contribution to the process of moderation the DT generator is not included in the actual simulation.}
\label{fig:portgeo}
\end{figure}

Simulations of viable shielding geometries were performed with a Geant4 based Monte Carlo application~\cite{allison2016_geant4,ref:grasshopper}.  Two series of simulations were performed to optimize both the moderator thickness and shielding geometries for the compact source.  All simulations used a 14.1 MeV isotropic neutron source to simulate neutrons produced by a compact DT source.  Neutron generation was modeled as an instantaneous pulse.   At later stages in the optimization analysis the time distribution of the output was smeared with a square function of a width 5~$\mu$s, which is the shortest possible pulse width of the commercially available Thermo Electron P325 neutron generator~\cite{ref:p325}.  In order to increase computational efficiency, a dedicated and separate simulation was performed to determine the optimal moderator thickness.  
The shielding optimization was performed by simulating a realistic experimental geometry, composed of: a concrete room filled with ambient air; the DT source shielding, collimation, and moderation; a tungsten-silver target as a test object; and an idealized detector with 100\% detection efficiency, surrounded by its own shielding to reduce room return.  Since the optimal moderator thickness was determined in the separate epithermalization simulations (see Section~\ref{sec:moderator_optimization}), 
these new simulations only iterated over the shielding and collimation geometries.  Moderator thicknesses were held constant at 10 cm for all geometries.  To suppress room return and collimator in-scattering, shielding geometry was modeled as a sphere of  75 cm radius consisting of of 5\% borated polyethylene.  This allowed to attenuate the off-axis neutron flux in order to reduce thermal and epithermal backgrounds.  To minimize the impact of in-scattering, the collimator walls and detector shielding was modeled to consist of boron carbide.  A schematic of the optimized shielding geometry is shown in Figure \ref{fig:portgeo}.  The target was chosen to consist of a mix of tungsten, molybdenum, and silver. These elements have very clear resonances in the $\sim$eV range, and thus can be used for an optimization analysis applicable to most actinides (e.g. $^{235}$U, $^{238}$U, and most plutonium isotopes).  Also, both tungsten and silver are easily available in the necessary quantities for performing follow-up experimental studies.  The shielding design included cadmium filters in the beam path, as a way of eliminating the thermal flux exiting the moderator and impinging the detector,  via the $^{113}$Cd(n,$\gamma)^{114}$Cd reaction "threshold" at 0.5 eV. See Table \ref{table:parameters} for the dimensions of the system model.

\subsection{Moderator Optimization} \label{sec:moderator_optimization}

\begin{figure}[h]
\centering\includegraphics[angle=270,width=0.8\linewidth]{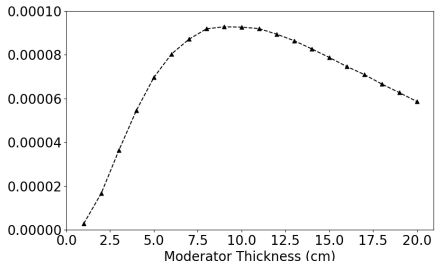}
\caption{Epithermalization efficiency of various moderator thicknesses.  The maximum efficiency occurs at approximately 9-10 cm with a stable range between 7 and 11 cm.  Competing processes of epithermalization and thermalization lead to the local maximum.}
\label{fig:efficiencycurve}
\end{figure}

The process of neutron moderation introduces uncertainty into the TOF reconstruction.  An optimal moderator maximizes the epithermalization efficiency and minimizes the spread in TOF.  The epithermalization efficiency was defined as the fraction of incident 14.1 MeV neutrons that were slowed down into the energy range 1-30 eV and exited the rear surface of the moderator.  Figure \ref{fig:efficiencycurve} shows a plot of overall epithermalization efficiency vs. moderator thickness.  The ideal moderator thickness was found to be approximately 10 cm, with minimal change between 7 and 11 cm.  The initial increase in efficiency with increasing moderator thickness is primarily a result of additional scattering of 14.1 MeV neutrons.  The decrease in efficiency beyond 11 cm occurs as down-scattering from the epithermal to thermal ranges begins to dominate. The maximum epithermalization efficiency was determined to be $(9.3\pm 0.1)\times 10^{-5}$.  Even for the optimized moderator thickness, the thermal flux at the rear surface of the moderator is two orders of magnitude higher than the epithermal flux since increasing cross sections at lower energies lead to a build-up of thermal neutrons in the system.  Most of these thermal neutrons can be removed using the cadmium filters.

The simulation output was also analyzed to understand the effect of moderation on the TOF energy reconstruction.  This analysis showed that changes in moderator thickness had little effect on the accuracy of TOF reconstruction in the epithermal regime in the range of flight distances considered.  Studies of moderation times indicates that, depending on energy, the spread in times are in the range of 0.1-1 \textmu s.  Thus, using appropriate digitizer or time-to-digital coincidence electronics with the time resolution of $\ll$ \textmu s is necessary. CAEN V1720 digitizers, as an example, have sampling at the period of 4 ns.  Such sampling will have an insignificant effect on TOF resolution.

\subsection{System Geometry}

The configuration described in Figure~\ref{fig:portgeo} succeeded in both minimizing the impacts of room return and in-scattering on the transmission spectrum.  Designs that incorporated less High Density Polyethylene (HDPE) or eliminated the shell suffered significant leakage from the outer surface of the absorber-moderator sphere, resulting in high intensities of room return and epithermal background.  We thus omit these configurations from this discussion.

\begin{table}[h]
\centering
\begin{tabular}{l l}
\hline
\hline
\textbf{Parameter} & \textbf{Dimension} \\
\hline
Distance from DT generator to moderator surface & 5 cm \\
Moderator thickness & 10 cm \\
Distance from surface of moderator to detector & 1.845 m \\
Collimator diameter & 20 cm \\
Outer radius of borated polyethylene  & 75 cm \\
Thickness of boron carbide plating & 1 cm \\
Thickness of detector shielding & 9 cm \\
Length of detector shielding  & 80 cm \\
\hline
\hline
\end{tabular}
\caption{Key dimensions of portable, epithermal DT source. Neutron flight path can be increased or decreased depending on needs of measurement.}
\label{table:parameters}
\end{table}

The detector collimator was also plated with boron carbide to minimize the probability that a neutron scattered off or through the surface of the collimator would be measured by the detector.  With this shielding, any neutron that does not enter the collimator is absorbed by the borated absorber-shield.  These off-axis neutrons have a negligible effect on the room return background of the system, as can be seen in the discussion below. 
The geometric model of the detector collimator/shield played an important role in the accuracy of the TOF reconstructions.  While no epithermal or thermal neutrons scattered from the concrete walls were observed to reach the detector, a simple cylindrical collimator geometry resulted in a significant number of detected epithermal neutrons which underwent in-scatter from the inner walls of the collimator.  For this simple geometry the line-of-sight (LoS) neutrons, i.e. neutrons that did not undergo inscatter and reached the detector directly from the source constituted only 85\% of the total.  By comparison when using a polycone geometry, described schematically in Fig.~\ref{fig:portgeo}, this number was increased to $(95.4 \pm 0.2)$\%, thus significantly improving the accuracy of the TOF technique.  The fraction of LoS neutrons can be further improved via future optimizations of the inner surfaces of the shielding.


\subsection{Target} \label{sec:target}
Simulated targets consisted of metallic cylinders 10 cm in diameter and between 5 and 10 mm in width.  Simulations were performed for five targets consisting of mixes of silver, molybdenum, and tungsten with varied molar ratios.  Mixes of silver and tungsten were simulated at 10\%, 20\%, and 30\% silver molar abundance.  Mixes of molybdenum and tungsten were simulated at 50\% and 75\% molybdenum molar abundance.  For brevity, isotopic compositions will be referenced using their composite elements followed by their atomic percent.  Using this convention, the mixes are referenced as Ag10:W90, Ag20:W80, Ag30:W70, Mo50:W50, and Mo75:W25, respectively. All materials were simulated with their natural isotopic abundances.

\section{Results} \label{sec:results}
The accuracy of the energy reconstruction and the isotopic sensitivity were both characterized using a $\chi^2$ test.  The TOF energy reconstructions were determined for the energy range of 1-15 eV using a histogram bin width of 0.1 eV.  
The goals of this study were to characterize the following:
\begin{enumerate}
    \item Accuracy of spectral reconstruction - determined by comparing the spectra of the TOF energy reconstruction and the actual neutron energy in an idealized detector.  Lower $\chi^2$ values indicated better reconstruction with an ideal reconstruction having a $\chi^2$ value of zero.
    \item Isotopic sensitivity - determined by directly comparing the TOF energy reconstructions for targets of differing isotopic composition.  Higher $\chi^2$ values indicated higher sensitivity.
\end{enumerate}


\subsection{TOF Energy Reconstruction}

In all MC simulations, the distance from the surface of the moderator to the detector was set to 1.845~m as a balance between TOF reconstruction precision and geometrical efficiency.  TOF energy calculations used a flight path length of 1.87~m because it resulted in more accurate reconstructions.  Further analysis confirmed that the mean flight path for incident LoS epithermal neutrons was approximately 1.87 meters, since most epithermal neutrons are epithermalized between 2 and 3 cm before the rear surface of the moderator.

\begin{figure}[h]
\centering
\includegraphics[angle=0,width=1.02\linewidth]{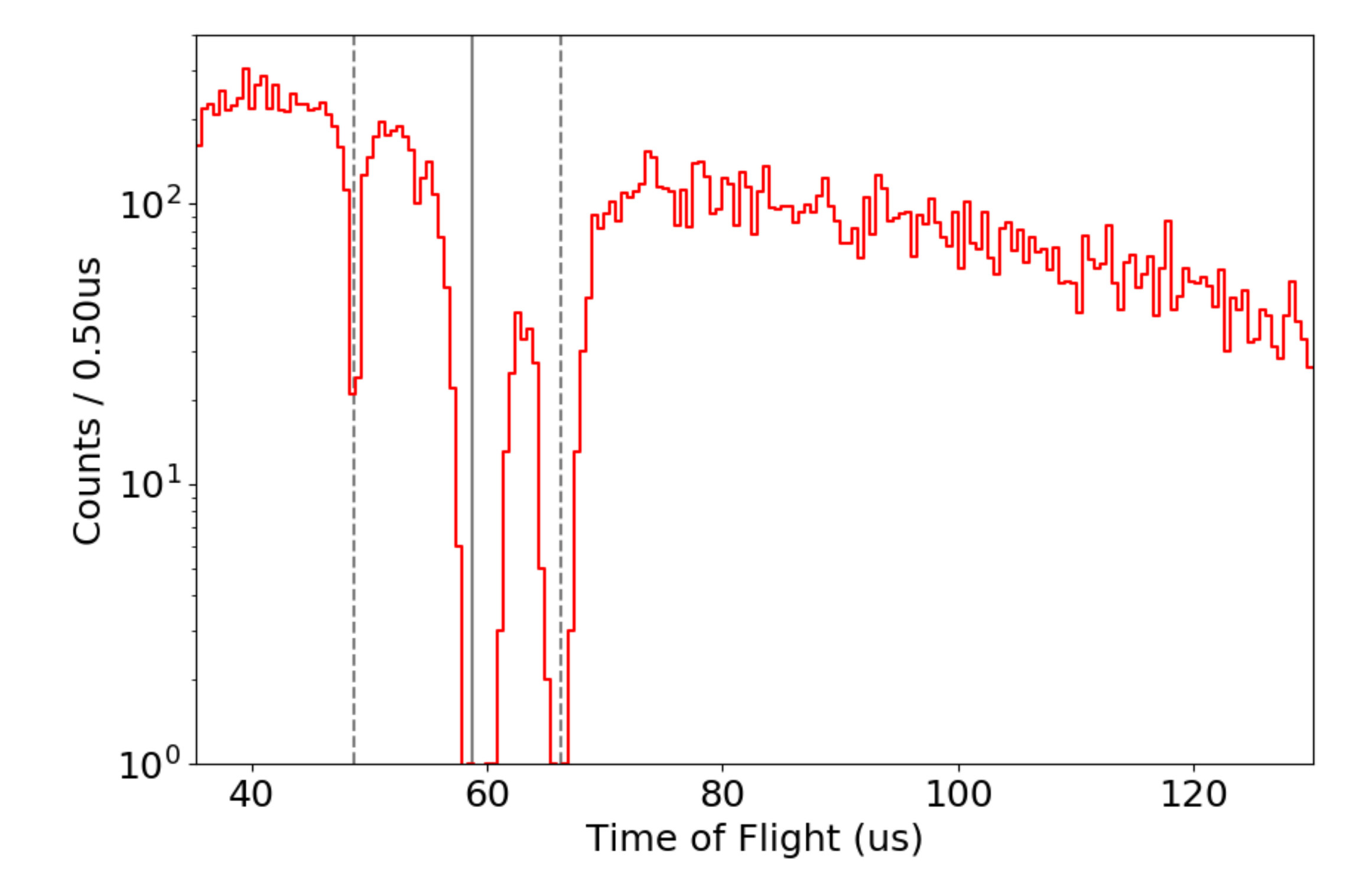}
\includegraphics[width=\linewidth]{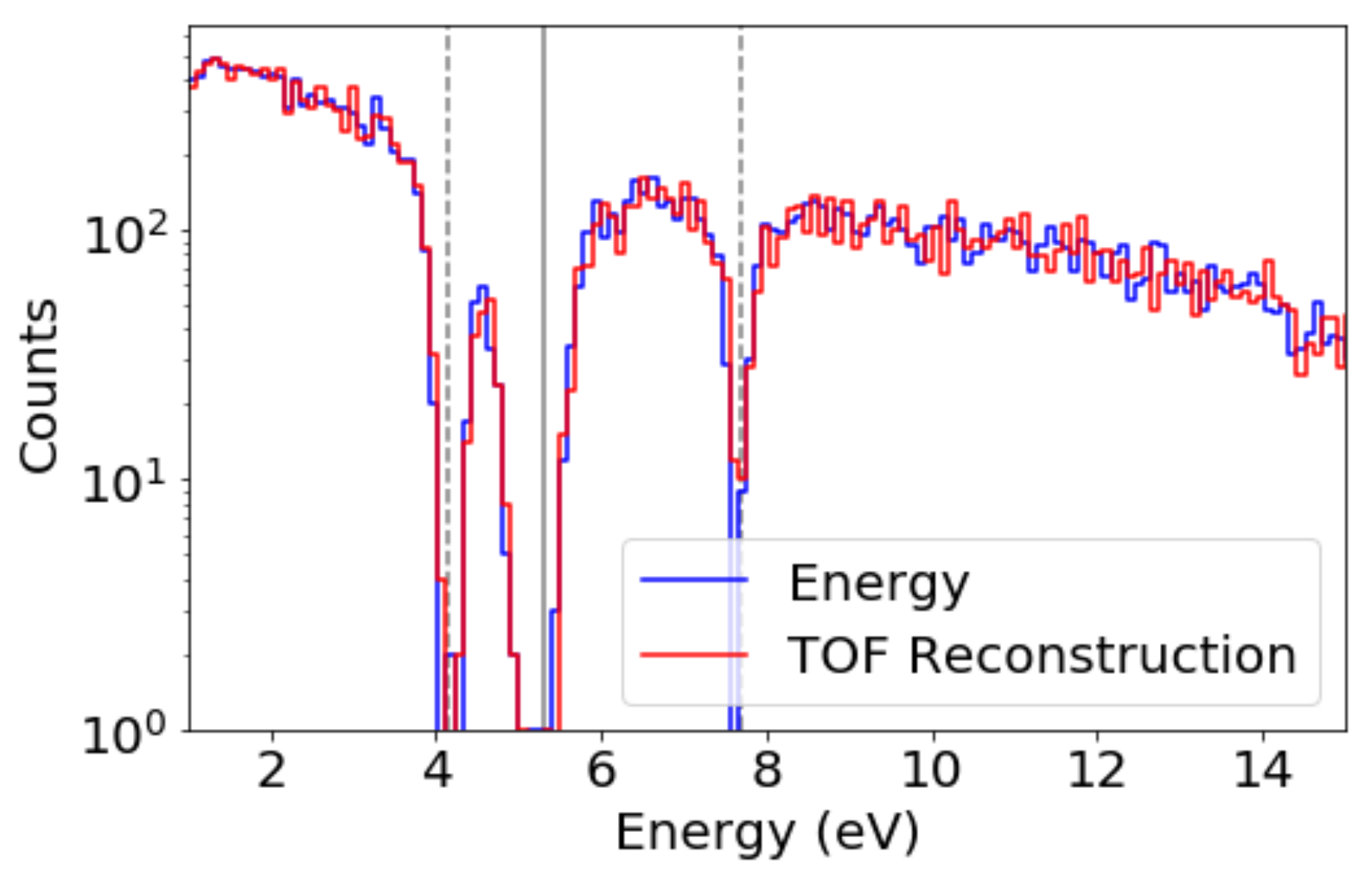}
\caption{Top:  TOF distribution.  Bottom: actual energy and TOF-reconstructed energy comparison. The simulation is for a Ag10:W90 target.  Dashed and solid gray lines denote the expected locations for the tungsten and silver resonances, respectively.}
\label{fig:TOFrec}
\end{figure}

Figure \ref{fig:TOFrec} shows a plot of the energy spectrum of the detected neutrons and the TOF reconstruction for a silver-tungsten mix (at a 10\% silver molar abundance).  The absorption lines due to the silver resonance (5.2 eV) and tungsten resonances (4.1, 7.7 eV) are clearly observed in both the actual and TOF-reconstructed energy spectra.  Table \ref{table:tof-rec} shows the $\chi^2$ for comparisons for various targets.  The value of $\chi^2/$DoF does not exceed unity for any of the materials or isotopic compositions indicating a reasonably good, but not ideal, energy reconstruction.

\begin{table}[h]
\centering
\begin{tabular}{l l l}
\hline
Target & \makecell[l]{Target thicknesses\\(mm)} & $\chi^2/$DoF \\
\hline
Ag10:W90 & 5 & 73 / 132 \\
Ag25:W75 & 5 & 39 / 129 \\
Ag30:W70 & 5 & 39 / 133 \\
Mo50:W50 & 10 & 52 / 135 \\
Mo75:W25 & 10 & 70 / 139 \\
\hline
\end{tabular}
\caption{$\chi^2$ values for comparison of actual energy to TOF-reconstructed energy for various targets.  The reconstructed energy spectra for all targets gives a $\chi^2/$DoF below 1.0, demonstrating reasonably accurate energy reconstruction.  The inaccuracy of the reconstruction is primarily caused by the smearing of the neutron pulse time distribution due to the process of moderation.  The DoF changes for different spectral comparisons due to the exclusion of bins with zero counts.}
\label{table:tof-rec}
\end{table}

\subsection{Isotopic Sensitivity}
One of the main advantages of NRTA is its ability to identify an object's material composition and to enable isotopic-geometric comparisons between two objects, such as in a warhead verification exercise~\cite{ref:hecla2018nuclear,ref:engel2019}.  The system's performance in the latter scenario is used as the basis for its characterization.  The spectral transmission comparisons are performed between objects of different isotopic compositions, enabling characterization of the technique's material discrimination capability.

\begin{figure}[h]
    \centering
    \includegraphics[angle=270, width=\linewidth]{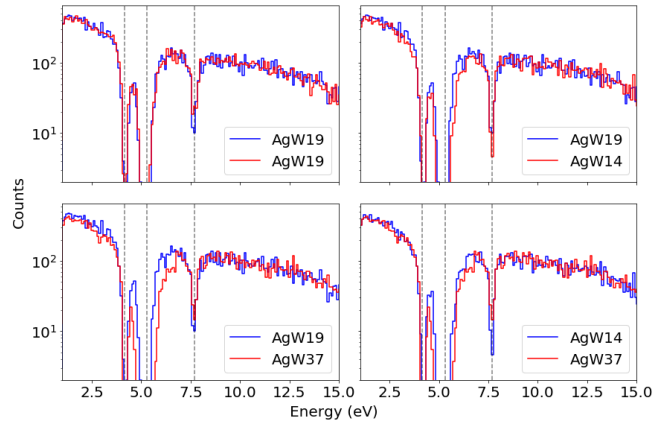}
    \caption{Spectral Comparisons of Various Targets.  Comparisons of TOF energy reconstructions for transmitted neutrons through different targets are presented. Visual inspection shows clear differentiation between varied isotopic mixes.  Dashed lines represent tungsten and silver resonances.  Table~\ref{table:spectra-sep} lists the results of the $\chi^2$ tests for these comparisons, showing a clear quantitative difference.}
    \label{fig:spectracomp}
\end{figure}

The isotopic sensitivity of the system was characterized by comparing the TOF reconstructions of varied materials (Ag10:W90, Ag20:W80, Ag30:W70, Mo50:W50, and Mo75:W25).  Molar abundances and thicknesses were chosen to optimize transmission probabilities and resonance reconstruction, as well as to allow for practical follow-up experiments.  Figure \ref{fig:spectracomp} shows the plots of spectral comparisons for a subset of test scenarios.  To evaluate the system's ability to distinguish between targets of different isotopic composition, the transmitted spectra were compared using a $\chi^2$ test.  Table \ref{table:spectra-sep} shows the $\chi^2/$DoF for the TOF-based spectral comparisons of various isotopic configurations.  A comparison of two simulations for the 10\%/90\% Silver-Tungsten target (with different random seeds) was used as a control.  Materials of various isotopic composition were compared with each other directly.  In the last two columns of the table we also present the results of simulations where the neutron pulse was smeared with a 5 $\mu$s square function, to emulate the typical duty of the most commonly available neutron generators, such as P325~\cite{ref:p325}.  All simulation results were scaled to $(1.9\pm0.3)\times10^{10}$ total 14.1 MeV neutron generation events, enabling direct comparison.  The comparisons show a clear differentiation between objects of different isotopic composition, even when the neutron pulse is smeared by 5 $\mu$s.  For a typical DT source, the above sample population would correspond to a total measurement time of approximately 30 minutes, assuming an isotropic source output of approximately $10^9$~neutrons/s.

\begin{table}[h]
\centering
\begin{tabular}{l l l l l l l}
\hline \hline
Material & Ratio1 &  Ratio2 & $\chi^2/$DoF & P-value & \makecell[l]{$\chi^2/$DoF \\ 5$\mu$s}&\makecell[l]{P-value\\ 5$\mu$s } \\
\hline
Ag-W & 10:90 & 10:90 & 102 / 85 & 0.10 & 82 / 85 & 0.57\\
Ag-W & 10:90 & 20:80 & 189 / 83 & $3\times10^{-10}$  & 152 / 85 & 0.00\\
Ag-W & 10:90 & 30:70 & 324 / 81 & $3\times10^{-30}$ & 301 / 83 & 0.00\\
Ag-W & 20:80 & 30:70 & 128 / 81 & $7\times10^{-4}$  & 113 / 82 & 0.01\\
Mo-W & 50:50 & 75:25 & 216 / 151 & $4\times10^{-4}$ & 225 / 154 & 0.00\\
\hline \hline
\end{tabular}
\caption{$\chi^2$ values for spectral separation analysis.  Column 1 denotes the isotopes present.  Columns 2 and 3 denote the molar ratios being compared.  Columns 4 and 5 denote the $\chi^2$ test statistic and the corresponding p-value for an instantaneous pulse and a detector at a distance of 1.87~m.  Columns 6 and 7 denote the same values for a DT pulse of 5$\mu$s and a detector placed at 5~m.  Separation increases monotonically as the molar abundance of silver in the second mix increases.  The control comparison has a $\chi^2$ statistic of approximately order unity.  The silver-tungsten and molybdenum-tungsten comparisons are restricted to energy ranges of 1.5-10 eV and 1.5-15 eV, respectively.}
\label{table:spectra-sep}
\end{table}

The control comparison (Ag10:W90 vs. Ag10:W90) has a reduced $\chi^2$ statistic of approximately order unity which is consistent with two samples drawn from the same distribution.  As the molar ratio of the silver is increased in the mix, the $\chi^2$ value compared to Ag10:W90 increases monotonically.  Figure \ref{fig:spectracomp} shows the spectra for the first four comparisons between mixed compositions of silver-tungsten.  The differences between the reconstructions are clearly visible in the spectra.  The separation between the 20\% and 30\% samples deteriorates due to limited statistics within the silver and tungsten resonance structures.  This is reflected by the relatively low $\chi^2$ value.  The Mo50:W50 and Mo75:W25 separation is also limited, and this is due primarily to low statistics within the resonances combined with a similar overall cross section for both configurations.  In both cases however, the spectra can be clearly differentiated.  In all of these simulations $2.5\times10^9$ neutrons were generated.  With a source that produces neutrons at the rate of $10^9$s$^{-1}$, individual runs would require 25 seconds.  Thus the compact configuration described in this study can distinguish between different isotopic compositions with total measurement times of approximately one minute. This time could be further reduced as new, more intense sources of neutrons become available. 

\subsection{Effects of Neutron Generation Pulse Width} \label{sec:pulsewidth}

\begin{figure}[h]
    \centering
    \includegraphics[width=\linewidth]{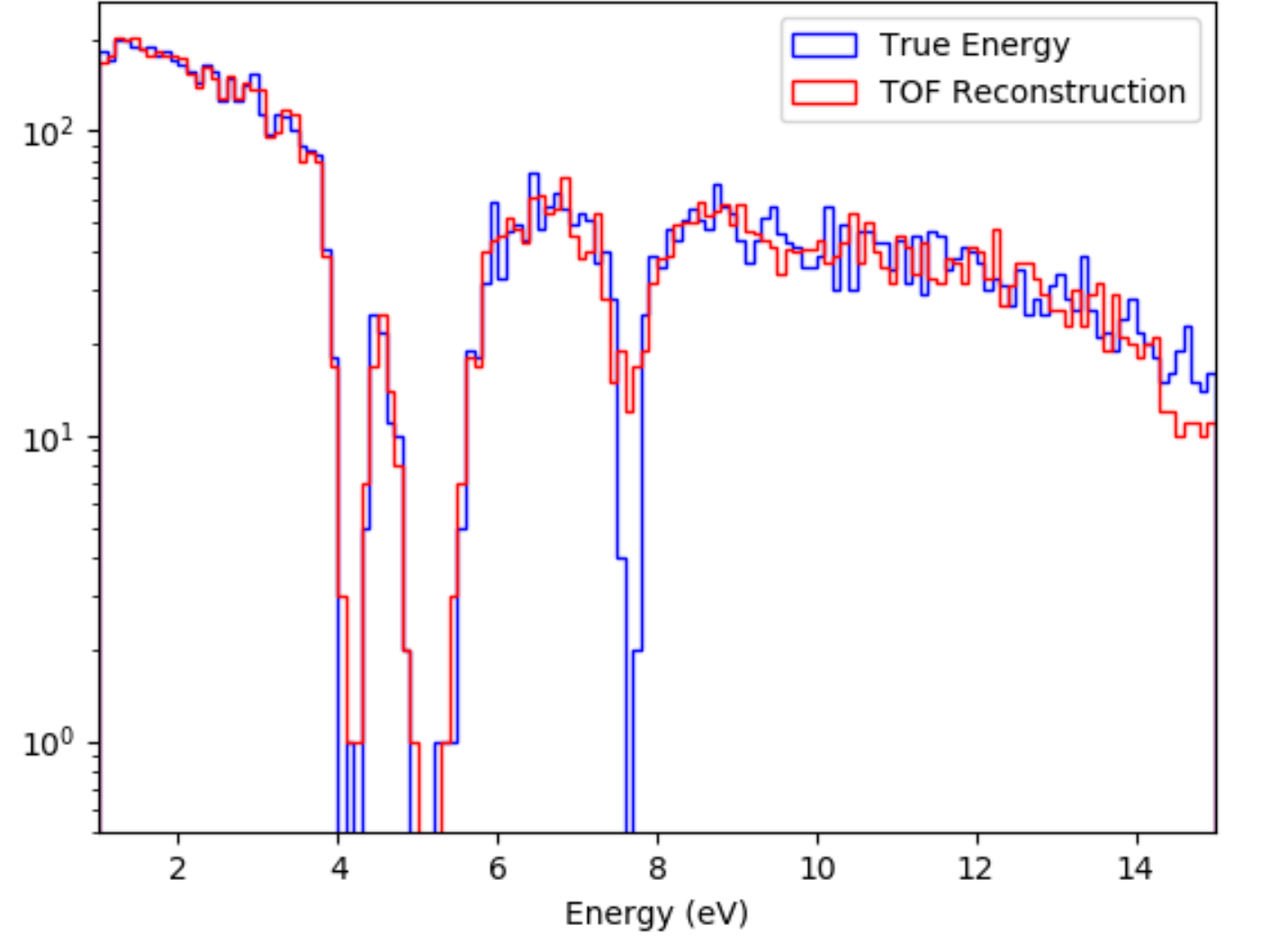}
    \caption{Effect of pulse width on TOF Reconstruction Quality. Comparison of Ag10:W90 transmission spectra for an instantaneous neutron generation pulse (blue) and a 5.0 $\mu$s pulse width (red) at 5.0~m TOF distance. Note the slight smearing of the narrow resonance at 7.7 eV and the difference in spectra at higher energy.  These effects are not seen in the spectra in Figures \ref{fig:TOFrec} and \ref{fig:spectracomp} produced without a finite pulse width.}
    \label{fig:pulsewidthspectrum}
\end{figure}

Up to this point, it was assumed that the neutron source generates a neutron beam instantaneously. While fast, sub-microsecond DT systems exist, the most commonly commercially available sources typically produce a $\sim5$ $\mu$s pulse.  To account for this circumstance the output of the simulation was modified by adding a uniformly random 0-5 $\mu$s offset to the TOF of every neutron event.  This pulse time introduces uncertainty into the $t_0$ of any given neutron, and as a result introduces an energy-dependent error in the TOF-reconstructed neutron energy (see Figure \ref{fig:pulsewidthspectrum}).  As with error contributions from the moderation time, this spectral smearing effect is more prominent at higher energies where the time of flight is shorter. This effect could be mediated by moving the detector further from the source and thus increasing the time of flight.  For this analysis the detector was placed at a TOF distance of 5~meters, with an equivalent neutron flux incident on the detector.  The last two columns of Table \ref{table:spectra-sep} show the reduced $\chi^2$ values for the various targets for a time-smeared pulse.  While a significant and expected deterioration of the statistical separation is observed, all spectra can be distinguished from one another.  This indicates that while the system's analyzing power is reduced, it is still capable of performing material differentiation.  For a realistic system, the optimal TOF distance will depend on the isotopic composition and thickness of the targets. 


\section{Conclusions}

The proposed neutron resonance transmission analysis (NRTA) system is capable of highly accurate TOF energy reconstruction and is highly sensitive to changes in isotopic composition of the target.  Errors in TOF reconstruction are small and due primarily to pulse smearing within the moderator and in-scattering from the detector shielding.  The polycone design significantly reduces these effects.  The reconstruction might be further improved by increasing the flight path and further optimizing the polycone design. The sensitivity to isotopic composition makes this methodology a powerful tool for on-site measurements and characterizations of nuclear and non-nuclear materials alike.  With further work the system may be able to characterize and identify unknown samples of material with epithermal resonance structures.

The use of compact neutron sources has the potential to significantly expand the utility and applicability of NRTA in a variety of fields.  
While this study has focused on the optimization of DT-based sources alone, DD-based and linac-based photodisintegration sources may also provide viable methods for compact neutron generation.  Such sources may require less shielding and enable more precise TOF determination via narrower pulses.  For example, new generation of high intensity DD sources can achieve ${\lesssim}$2$\mu$s pulses~\cite{ref:starfire}, allowing for high TOF precision and significantly reduced shielding.  Furthermore, a 3 MeV linac can produce fast-pulsed neutrons in the $\sim$200 keV range, requiring minimal neutron shielding and collimation.  Future research should explore these alternatives.

Finally, this study did not take into the account the background contributions of gammas from neutron capture on hydrogen and boron.  Past work by \citeauthor{wang2016improved}\cite{wang2016improved} using $^6$Li glass detectors have indicated that suppression of gamma counts by $\mathcal{O}(10^6)$ is feasible.  Furthermore, the pulsed mode of the source operation can allow for further reduction of gamma backgrounds: as capture gammas come primarily from thermal neutron capture their arrival times are more spread out in time than that of the 1-20 eV neutrons.  Nevertheless, the careful choice of the detector and the data analysis algorithms is important for controlling the background and should be part of future experimental efforts.

\section*{Acknowledgements}

This work is supported in part by Department of Energy Award DE-NA0002534, as part of the NNSA Consortium of Verification Technologies (CVT). The authors would like to thank their colleagues in CVT and MIT's Nuclear Science and Engineering for encouragement and advice.


\bibliography{bibliography}

\end{document}